\documentclass[11pt]{article}
\newcommand{\ba}{\begin{array}}
\newcommand{\bc}{\begin{center}}
\newcommand{\be}{\begin{equation}}
\newcommand{\ber}{\begin{eqnarray}}

\newcommand{\bt}{\beta}

\newcommand{\ea}{\end{array}}
\newcommand{\ear}{\end{eqnarray}}
\newcommand{\ec}{\end{center}}
\newcommand{\ee}{\end{equation}}

\newcommand{\fr}{\frac}

\newcommand{\implies}{\Rightarrow}

\newcommand{\La}{\Lambda}

\newcommand{\lb}{\label}

\newcommand{\Lg}{{\cal L}}

\newcommand{\Om}{\Omega}

\newcommand{\p}{\partial}
\newcommand{\ph}{\phi}

\newcommand{\Si}{\Sigma}

\newcommand{\ta}{\tau}

\newcommand{\te}{\theta}

\begin{document}
\title{Galactic Metrics.}
\author{Mark D. Roberts,
54  Grantley Avenue,  Wonersh Park GU5 OQN,\\
Email: mdrobertsza@yahoo.co.uk,  
http://cosmology.mth.uct.ac.za/$\sim$~roberts}
\maketitle
\bc Comments:  7 pages. {\tt astro-ph/0209456}\ec
\begin{abstract}
The paths of stars in galaxies have circular velocity independent 
of their distance from the centre of the galaxy.
Newtonian mechanics with a logarithmic potential has such paths.
In relativity these paths can be taken to be geodesics and this requirement places a
resticting equation on the spacetime metric.   
This equation has a non-unique spherically symmetric solution
that in the newtonian limit has a logarithmic potential.
It also can be solved in terms of a conformal factor.
In particular it has solutions which are conformal 
to the vacuum-Einstein solutions
and such spacetimes are solutions to the vacuum-Bach equations.
Therefore it is suggested that the Bach equations 
describe dynamics on galactic lenght scales.
\end{abstract}
\section{Introduction}\lb{intro}
The majority of galaxies conform to a few stereotyped structures such
as spirals and barred spirals.  The speed at which stars rotate around the
centre of such galaxies is not what would expected from their detected mass 
- rather the rotation speed is independent of the distance of the star from 
the centre of the galaxy.   Both optical and radio observations support this 
conclusion.   The usual solution to this problem is to invoke the existence 
of a large amount of as yet undetected material which resides in contrived 
places.   This is the so-called the ''{\sc missing mass}'' problem.   
Modern relativistic theories of gravitation equate the effects of matter 
with curvature.   Curvature is described by the Riemann tensor which contains 
derivatives and products of the Christoffel symbol.   The Christoffel symbol
is constructed from derivatives of the metric,  
which is an indicator of the distance between points.
In \S2 it is shown that in
Newtonian kinematics,  constant orbital velocity requires a potential of $\ln(r)$;
this result was mentioned without derivation in \cite{mdr91}.
The logarthimic potential is related to the metric and gives Christoffel 
symbols with terms proportional to $1/r$.   
The products and derivatives of these are both proportional to $1/r^2$;  
the standard approximation scheme uses only the highest derivative terms 
and hence production of the linearized Riemann tensor breaks down.  
Thus it is impossible to establish contact between 
linearized relativistic theory and Newtonian theory solely from knowledge
of a logarithmic potential.   
The relativistic treatment has to be approached in a different manner:   
the method that I suggest is to note that the paths of particles 
are given by geodesics which are curves of minimum length.   
For galaxies the geodesics are known - 
they are simply the circulating stars' rotation curves,  
called {\it galactic rotation curves}.   These have recently been reviewed in
detail by Blattaner and Florido \cite{BF} (2000).
Once the geodesics are known it is possible to find a coordinate dependent
constraint on the metric of the spacetime that the geodesics inhabit;
how this is done is described in \S3.

Observations of interaction between galaxies suggest that even more missing
mass is required,  and in cosmology large amounts of missing mass are also 
postulated:  the pattern is the longer the length scale the higher the 
proportion of hypothetical undetected matter.   
There is a critical density above which the Universe is closed and will contract.  
Nearly all cosmologists feel that 
the Universe is close to this critical density.   Using Einstein's equations 
this critical density requires that about 90\% of the Universe is as yet 
undetected.   An alternative is that different equations might be appropriate,
again the new equations originating from a relativistic theory other than 
Einstein's.   It would be expected that the appropriate equations will be the 
same ones that have a solution with a metric which explains constant galactic 
rotation curves.   

There are two main sorts of explanation for constant rotation curves.
Currently the most widely accepted is that there is a large amount of as yet
undetected mass,  see Blattner and Florido \cite{BF} (2001) and references therein.
The second is that force laws other than Newton's are needed,  
see in addition to \cite{BF},  
Roberts \cite{mdr1985} (1985) and \cite{mdr91} (1991) and references therein.
There are also explanations built on analogies with other systems,
such as hurricanes,  see Emanuel \cite{emanuel} (1991).
The explanation suggested here is that galactic spacetime 
is a solution to the Bach \cite{bach} (1921) quadratic equations.
Quadratic lagrangians can be expressed as two independent parts:
a $R^2$ short-ranged part and a conformally invariant $C^2$ long-ranged part,  so that
for present purposes only the $C^2$ part is of interest.
Specifically varing the $C^2$ action one obtains the Bach tensor
\be
B_{ab}=2C_{a..b}^{~cd}R_{cd}+4C_{a..b;cd}^{~~cd}.
\lb{bach}
\ee
Any spacetime with vanishing Weyl tensor $C_{abcd}=0$,  such as $R_{ab}=\La g_{ab}$, 
is a solution to the vacuum-Bach equations \ref{bach};
in particular any conformally flat spacetime 
and also K\"ottler's solution
\be
ds^2=-\left(1-\fr{2m}{r}+\fr{\La r^2}{3}\right)dt^2
     +\left(1-\fr{2m}{r}+\fr{\La r^2}{3}\right)^{-1}dr^2
     +r^2d\Si_2^2,
\lb{kotle}
\ee
are solutions.
This suggests that the large-scale vacuum,  
see the conclusion of \cite{mdrvac},  can be thought of as having 
a conformally flat metric rather than a flat one.
One is left with the situation that large scale dynamics are described by the 
Einstein-Bach equations,  the Einstein part dominating on lenght scales of the solar system 
and the Bach part dominating on galactic lenght scales.
This suggests that on cosmological scales 
field equations corresponding to quadratic lagrangians provide 
a better description than using the Einstein field equations:
cosmology and quadratic lagrangians have been reviewed 
in Querella \cite{querella} (1999) and references therein.
How constant rotation curves constrain axi-symmetric spacetimes 
has been discussed in Roberts \cite{mdr86} (1986).
What happens as $r\rightarrow\infty$ is a problem
in general \cite{mdrse} and particularly with any explanation
of constant rotation curves.
A solution is to take the spacetime solution to be the correct model 
for what was previously taken to be the galactic halo,
and this should be used for models of gravitational lensing.
If one takes the Bach equations to be the correct ones,  one simply
assumes that for sufficiently large $r$ a different solution applies.
The situation is similar to stars where one can take 
a different Schwarzschild exterior solution for each star,
for galaxies one takes a different anti-deSitter solution for each galaxy;
a difference with this analogy is it is possible to find the size
of the coupling constant by linearization for the Schwarzschild solution,
but unfortunately for galaxies there is as yet 
no known linearization to fix any of the quadratic coupling constants.
Constant rotation curves by themselves
do not explain the Tully-Fischer \cite{bi:TF} (1977) 
luminosity to speed relationship,
but perhaps there is a metric which explains both.
It is surprising that spherically symmetric spacetimes 
descibe constant rotation curves at all,
galaxies rotate and so are better modeled by axi-symmetric spacetimes.

The conventions used are those of \cite{HE},
in particular the signature is $(-,+,+,+)$.
Here all work is done in the plane,  so that $\dot{\te}=0$,
the reason that this can be done without loss of generality for 
spherically symmetric spacetimes are the same as those given in
Chandrasekhar \cite{cha} (1983).
\section{Newtonian Analysis.}\lb{newt}
In this section
it is shown that in Newtonian theory a logarithmic potential
gives circular paths independent of $r$.
In spherical coordinates Newton's second law is
\be
V_a/m=(V_r,V_\ph)/m=(\ddot{r}-r\dot{\ph}^2,\fr{d}{dt}r^2\dot{\ph}),
\lb{2pl}
\ee
by spherical symmetry the $\ph$ component integrates immediately to give
\be
r^2\dot{\ph}=L.
\lb{eqL}
\ee
Subsituting using $u\equiv 1/r$ and using $u'\equiv du/d\ph$ and \ref{eqL}, gives
\be
\dot{r}=-Lu',~~~
\ddot{r}=-L^2u^2u".
\lb{rL}
\ee
Using \ref{rL} the velocity vector is
\be
\underline{v}=\left(\dot{r},\fr{\dot{\ph}}{r^2}\right)=L(-u',1),
\lb{Lte}
\ee
and also using \ref{rL}
the first component of Newtons second law \ref{2pl} in polar coordinates is
\be
\fr{d^2u}{d\ph^2}+u=-\fr{V_r}{mL^2u^2}
\lb{bin}
\ee
which is called the Binet equation:
an imporatnt thing to note is that there is an undifferentiated $u$ on the LHS.
Recalling $g^{\ph\ph}=1/r^2$;
the square of the velocity at any point is
\be
v^2=\dot{r}^2+(r\dot{\ph})^2
=L^2(u'^2+u^2),
\lb{11}
\ee
for circular orbits $u'=0$ so that this is 
\be
v_c^2=r^2\dot{\ph}^2=\fr{L^2}{r^2},
\lb{eq5}
\ee
for circular orbits $u"=0$ so that from \ref{bin}
\be
L^2=-\fr{V_r}{mu^3}=-\fr{r^3}{m}V_r,
\lb{6}
\ee
substituting in \ref{eq5}
\be
v_c^2=\fr{L^2}{r^2}=-\fr{r}{m}V_r.
\lb{eq7}
\ee
For constant circular orbits 
\be
mv_c^2=-rV_r,
\lb{eq8}
\ee
or
\be
V=-mv_c^2\ln(r),
\lb{9}
\ee
the desired result.
The velocity and force vector fields are
\be
\underline{v}=(0,L)=(0,rv_c),~~~~~
\fr{V_a}{m}=\left(-\fr{v_c^2}{r},0\right).
\lb{2vec}
\ee
To see that the correct rotational velocity is $g^{\ph\ph}v_c^2=v^2_c/r^2$, 
rather than just $v_\ph^2$;  
note that using $v^2_\ph$ one gets
\be
v_\ph^2=L^2,~~~
V=\fr{mv_\ph^2}{2r^2},
\lb{19}
\ee
for \ref{eq5} and \ref{9} respectively,  the potential 
is shorter ranged potential than that of Newtonian gravitation.
\section{Spherically Symmetric Geodesics.}\lb{ssg}
Here geodesics are derived by the Lagrange method,
as used for example in Chandrasekhar \cite{cha} (1983).
The lagrangian for geodesics is 
\be
2\Lg=g_{tt}\dot{t}^2+g_{\ph\ph}\dot{\ph}^2+g_{rr}\dot{r}^2+g_{\te\te}\dot{\te}^2=p_ap^a,
\lb{s1}
\ee
in the $(-,+,+,+)$ convention, $2\Lg=-1$ for timelike geodesics.
The Euler equations are
\be
\fr{dp_a}{d\ta}=\fr{\p \Lg}{\p x^a}.
\lb{s2}
\ee
For \ref{s1}
\be
p_a\equiv\fr{\p\Lg}{\p\dot{x}^a}
=(g_{tt}\dot{t},g_{\ph\ph}\dot{\ph},g_{rr}\dot{r},g_{\te\te}\dot{\te}),~~~~~
p^a=(\dot{t},\dot{\ph},\dot{r},\dot{\te}),~~~~~
p_ap^a=-1,
\lb{s3}
\ee
and
\be
2\fr{\p\Lg}{\p x^a}=
(0,0,g_{tt,r}\dot{t}^2+g_{\ph\ph,r}\dot{\ph}^2+g_{rr,r}\dot{r}^2+g_{\te\te,r}\dot{\te}^2,0).
\lb{s4}
\ee
The first two components of the Euler equation integrate to give
momentum and its derivatives
\be
p_a=(E,L,g_{rr}\dot{r},g_{\te\te}\dot{\te}),
\lb{s5}
\ee
where $E$ and $L$ are constants for each geodesic.  
For the second two components
\be
\fr{dp_r}{d\ta}=g_{rr}\ddot{r}+g_{rr,r}\dot{r}^2,~~~
\fr{dp_\te}{d\ta}=0=\fr{d}{d\ta}g_{\te\te}\dot{\te},
\lb{s6}
\ee
working in the plane $\dot{\te}=0$ so the last of these is satisfied,
from \ref{s4} and \ref{s6} the $\dot{r}$ component of the Euler equation gives
\be
0=2g_{rr}\ddot{r}+g_{rr,r}\dot{r}^2-g_{tt,r}\dot{t}^2-g_{\ph\ph,r}\dot{\ph}^2,
\lb{g7}
\ee
in terms of $u"$, $\ddot{r}$ takes exactly the same form as in the Newtonian case \ref{rL},
as does $\dot{\ph}$ in equation \ref{eqL}.
Substituting for $\dot{t}$ using \ref{s6} then
dividing \ref{g7} by $-2L^2u^2$ 
and taking $g_{\ph\ph}\sim r^2$,  this can be seen to be a generalization
of \ref{bin} the Newtonian Binet equation
\be
0=g_{rr}u"-g_{rr,r}\fr{u'^2}{2u^2}+u+\fr{g_{tt,r}}{2}\left(\fr{E}{uLg_{tt}}\right)^2.
\lb{gnb}
\ee

The dynamics of geodesics in the plane are described by equations \ref{s1} and \ref{g7}.
For circular orbits again $u"=u'=0\implies\ddot{r}=\dot{r}=0$,  
so that the two equations become
\be
0=g_{tt,r}\dot{t}^2+g_{\ph\ph,r}\dot{\ph}^2,~~~
0=g_{tt}\dot{t}^2+g_{\ph\ph}\dot{\ph}^2-2\Lg.
\lb{s10}
\ee
For non-null geodesics $2\Lg\ne0$,
then these equations \ref{s10} have solution
\be
\dot{t}^2=\fr{2\Lg}{h}g_{\ph\ph,r},~~~
\dot{\ph}^2=-\fr{2\Lg}{h}g_{tt,r},~~~
h\equiv g_{tt}g_{\ph\ph,r}-g_{\ph\ph}g_{tt,r}\ne0,
\lb{s12}
\ee
for null geodesics \ref{s10} implies $h=0$, so that these equations do not hold.
Defining the angular velocity as
\be
\Om\equiv\fr{d\ph}{dt}=\fr{d\ph/d\ta}{dt/d\ta}=\fr{\dot{\ph}}{\dot{t}}
=\fr{g_{tt}}{g_{\ph\ph}}\fr{L}{E},
\lb{g9}
\ee
so that the geodesic momentum vector \ref{s3} is
\be
p^a=(\dot{t},\dot{\ph})=\dot{t}(1,\Om)=\dot{t}(1,\fr{v_c}{r}).
\lb{g9b}
\ee
From \ref{s12} and \ref{g9} the differential equation governing angular motion is 
\be
g_{tt,r}+g_{\ph\ph,r}\Om^2=0.
\lb{g11}
\ee
For axi-symmetric spacetimes a much longer calculation shows that this generalizes to
\be
\Om^{-4}g_{tt,r}^2
\left(g_{tt,r}-2\Om g_{t\ph,r}+\Om^2g_{\ph\ph,r}\right)
\left(g_{tt,r}+2\Om g_{t\ph,r}+\Om^2g_{\ph\ph,r}\right)=0.
\lb{ag13}
\ee
For constant angular velocity,  integrating \ref{g11}
\be
g_{tt}=-\Om^2\int g_{\ph\ph,r}dr=k-\Om^2g_{\ph\ph},
\lb{g13}
\ee
taking the constant of integration $k=-1$ and $g_{\ph\ph}=r^2$
\be
g_{tt}=-(1+\Om^2r^2).
\lb{g14}
\ee
Note that for the signature $(+,-,-,-)$,  \ref{g11} remains the same,
and the constant of integration is $k=+1$.
\ref{g14} is of the same form as the de-Sitter metric \ref{kotle} with $\La=+3\Om^2$.
For constant rotational velocity \ref{eq5} becomes $v_c^2=r^2\Om^2$ so that
\be
r^2g_{tt,r}+g_{\ph\ph,r}v_c^2=0,
\lb{g15}
\ee
Integrating with $g_{\ph\ph}=r^2$ and,  as before,  choosing that constant of integration 
to give the correct flat spacetime limit gives
\be
ds^2=-\left(1+2v_c^2\ln(r)\right)dt^2+g_{rr}dr^2+r^2d\Si_2^2,
\lb{g16}
\ee
which agrees with the newtonian potential \ref{9}. 
The tensors for logarithmic metrics of this form suggest no easy exact solution.

This is not the unique solution of \ref{g15} by any means,
assuming a conformally flat form for the metric
\be
\bar{ds}^2=\Xi^2ds^2_M=\Xi^2\left(-dt^2+dr^2+r^2d\Si_2^2\right)
\lb{g17}
\ee
and after dividing by $2r\Xi$,  \ref{g15} becomes
\be
(v_c^2-1)r\Xi_r+v_c^2\Xi=0,
\lb{g18}
\ee
which has solution
\be
\Xi=Ar^{\fr{v_c^2}{1-v_c^2}}
\lb{g19}
\ee
where $A$ is a constant.   Defining the luminosity coordinate
\be
R\equiv\Xi r=Ar^{\fr{1}{1-v_c^2}}
\lb{g20}
\ee
the metric becomes
\be
ds^2=-A^{2(1-v_c^2)}R^{2v_c^2}dt^2+(1-v_c^2)^2dR^2+R^2d\Si_2^2,
\lb{g21}
\ee
as $r\rightarrow\infty,~ g_{tt}\rightarrow\infty$ 
so that the metric is not asympotically flat.
The metric \ref{g17} is conformally flat so that 
it is a solution to the vacuum Bach equations.
A generalization of \ref{g17} is to metrics conformal to known solutions,
c.f.eq.2.28\cite{HE}.
The generalization of \ref{g18}
\be
\fr{\Xi_r}{\Xi}=-\fr{v_c^2}{(v_c^2+g_{tt})r}-\fr{g_{tt,r}}{2(v_c^2+g_{tt})},
\lb{g22a}
\ee
which has particular solutions
\be
\Xi=Ar^\fr{v_c^2}{1-v_c^2}\left\{-v_c^2-g_{tt}\right\}^\bt
\lb{g22b}
\ee
where $\bt=1/(2(v_c^2-1))$ for deSitter spacetime 
and $\bt=(3v_c^2-1)/(2(1-v_c^2))$ for Schwarzschild spacetime 
and there is no $\bt$ for both $m,\La\ne0$.
The resulting line elements are not explicitly calculable 
in terms of the luminosity coordinate $R$.

Another approach to constant rotation curves 
is to assume a spherically symetric solution and use \ref{ag13}
to turn the spacetime an axially symmetric one with the correct geodesics.
The first bracketed term of \ref{ag13} integrates to
\be
g_{t\ph}=v_cr+\fr{1}{2v_c}\int g_{tt,r}rdr
        =v_cr-\fr{m}{v_c}\ln(r)+\fr{\La r^3}{9v_c},
\lb{g24}
\ee
[label g24]
where the integral is evaluated for K\"ottler's spacetime \ref{kotle}.
\section{Conclusion.}\lb{conc}
The problem of constant galactic rotation curves has been looked at in three ways.
The {\it first} suggested metric \ref{g16} agrees with the newtonian solution to the problem,
however its tensors appear to obey no simple field equations,
as there remains an arbitrary function $g_{rr}$ 
in the line element this might be rectified in the future.
The {\it second} is to produce spacetimes conformal \ref{g17} to known solutions,
these spacetimes obey the vacuum-Bach equations \ref{bach},
but in approximation do not give back the newtonian theory.
The {\it third} is to add an axi-symmetric component to known spherically symmetric solutions,
these spacetimes do not seem to obey known field equations or reduce to newtonian theory.
\section{Acknowledgement.}
I would like to thank Tom Kibble,  Mikhail Medvedev 
and Daniel Sudarsky for commenting on this paper.

\end{document}